\documentclass[12pt, preprint]{iopart}
\usepackage{graphicx}
\usepackage{color}
\begin{document}

\title[Self-excited current oscillations in a resonant tunneling diode]{Self-excited current oscillations in a resonant tunneling diode described by a model based on the Caldeira-Leggett Hamiltonian}
\author{Atsunori Sakurai$^1$ and Yoshitaka Tanimura$^{1,2}$}
\address{$^{1}$ Department of Chemistry, Graduate School of Science, Kyoto University, Kyoto 606-8502, Japan}
\address{$^{2}$ Department of Chemistry, Technical University of Munich, D-85747 Garching, Germany}\ead{sakurai@kuchem.kyoto-u.ac.jp and tanimura@kuchem.kyoto-u.ac.jp}
\begin{abstract}
The quantum dissipative dynamics of a tunneling process through double barrier structures is investigated on the basis of non-perturbative and non-Markovian 
treatment. We employ a Caldeira-Leggett Hamiltonian
with an effective potential calculated self-consistently, accounting for the electron distribution.
With this Hamiltonian, we use the reduced hierarchy equations of motion in the Wigner space representation to study non-Markovian and non-perturbative thermal effects at finite temperature in a rigorous manner.
We study current variation in time and
the current-voltage (I-V) relation of the resonant tunneling diode for several widths of the contact region,
which consists of doped GaAs.
Hysteresis and both single and double plateau-like behavior are observed in the negative differential resistance (NDR) region.
While all of the current oscillations decay in time in the NDR region in the case of a strong system-bath coupling,
there exist self-excited high-frequency current oscillations in some parts of the plateau in the NDR region in the case of weak coupling.
We find that the effective potential in the oscillating case possesses a basin-like form on the emitter side (emitter basin)
and that the current oscillation results from tunneling between the emitter basin and the quantum well in the barriers.
We find two distinct types of current oscillations, with large and small oscillation amplitudes, respectively.
%The results of eigenstates analysis indicate that the first type is caused by 
%a transition
%between ground tunneling states and adjacent excited states in the emitter basin,
%while the second type is caused by a transition between intermediate tunneling 
%states and higher states.
These two types of oscillation appear differently in the Wigner space,
with one exhibiting tornado-like motion
and the other exhibiting a two piston engine-like motion. 
%The I-V curves and the Wigner distribution obtained from the Boltzmann 
%equation approach are also presented and compared.
\end{abstract}

\maketitle

%////////////////////////////////////////////////////////////////////////

%========================================================================
\section{Introduction\label{sec:intro}}

Quantum coherence and its destruction by coupling to a dissipative
environment play an important role in the transport phenomena of a particle
moving in a potential \cite{CLAnnlPhys1983, Feynman63,Grabert88}.
Well known examples include electron transfer in molecular and biological systems \cite{Garg1985,Sparpagilione1988},
many chemical reactions \cite{Wolyness1981, WaxmanLeggett1985,Miller1989},
SQUID rings \cite{Chen1986, Wellstood2008},
quantum ratchets \cite{Hanggi97,Hanggi09},
nonlinear optical processes \cite{Mukamel95, TaniIshi09,TM2D1993,TMPRE1993,TOJCP1997},
and tunneling processes in device systems \cite{Datta, FerryGoodnickBird}.
Such systems are commonly modeled as one-dimensional or two-dimensional potential systems
coupled to heat bath degrees of freedom, which drive the systems
toward the thermal equilibrium state.
The heat bath degrees of freedom are then reduced using
such methods as the projection operator method or the path integral method, for example.
Many equations of motion have been derived for the purpose of understanding the quantum aspects
of dissipative dynamics \cite{Kadanoff1962,Mahan,Jauho2008,CLPhysica1983,Waxman1985, Cao1997, Cao1996,Burghardt2005,Burghardt2009, Martin2010, Martin2013}.

Because a complete picture of quantum dissipative
dynamics must treat phenomena that can only be described in real time,
a great deal of effort has been dedicated to the problem of numerically
integrating these equations of motion in real time \cite{Cao1997,Burghardt2005,Burghardt2009,Martin2010, Martin2013,Zueco04,Coffey07,Coffey07PCCP,Jyoti2011}.
Although these equations
are analogous to the classical kinetic equations,
which have proved to be useful for classical transport problems,
such equations cannot be derived in a quantum
mechanical framework without significant approximations and/or assumptions.
For example, the quantum Boltzmann equation is based on the assumption that the effects of collisions
between electrons can be described by the rates determined from Fermi's golden rule,
and hence it is regarded as a semi-classical equation \cite{Kadanoff1962,Mahan,Jauho2008}.
Similarly, the quantum Fokker-Planck equation can be derived from the Caldeira-Leggett
Hamiltonian under a Markovian approximation,
but in order for this to be possible,
the heat bath must be at a sufficiently high temperature,
in which case most of the important quantum dynamical effects play a minor role \cite{CLPhysica1983,Waxman1985}.
Treatments of these kinds are therefore not sufficient to
construct fully quantum mechanical descriptions of broad validity.
%In this paper, we present a rigorous quantum mechanical treatment that solves this problem.

To circumvent this problem, we present a quantum mechanical approach, which is valid for arbitrary temperatures.
This treatment employs the reduced hierarchy equations of motion (HEOM),
and it can be used in application to systems for which fully quantum mechanical description is necessary \cite{TJPSJ2006,TKJSPS1989}. 
In particular, the reduced HEOM approach can be used to numerically treat non-Markovian system-bath coupling
in a non-perturbative manner \cite{Tanimura91,Tanimura92, ITJSPS2005,TM97,Tanimura00,Kato02,Kato04,Ishi07,STJPCA2011,KTJPCB2013}.
Here, we apply this approach to study the dynamics of a resonant tunneling diode (RTD) described by the Caldeira-Leggett
Hamiltonian.

\begin{figure}
\begin{center}
  \includegraphics[width=130mm]{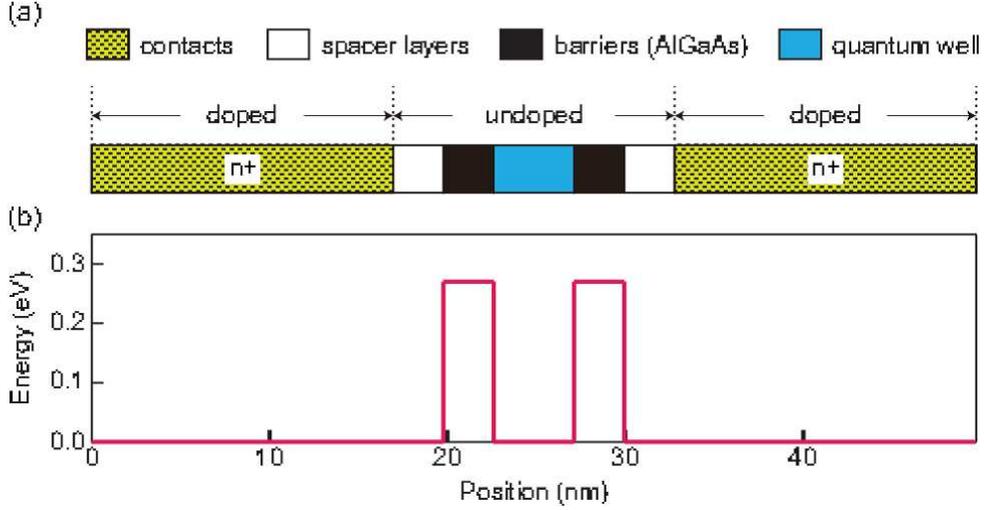}
\end{center}
\caption{\label{fig1}
(a) The structure of the RTD.
The well consists of undoped GaAs (4.520 nm),
the barriers consist of undoped AlGaAs (2.825 nm),
the spacer layers consist of undoped GaAs (2.825 nm),
and the contact regions consist of doped GaAs (with a doping concentration of $2\times10^{18}$ $\mathrm{cm}^{-3}$).
In order to elucidate the dependence of the current on the width of the contact regions,
we carried out computations for three values of this widths:
(i) 16.950 nm; (ii)  33.900 nm; (iii) 42.375 nm.
This figure depicts the case of 16.950 nm.
(b) The structure of the conduction band edge.
The height of the potential barriers is 0.27 eV.
}
\end{figure}

The RTD system that we consider is modeled by a double barrier structure with an electrostatic potential
representing a region consisting of an undoped layer positioned between two doped layers. (See figure. \ref{fig1}).
The double barrier structure constitutes a single quantum well with discretized energy states for electrons which are called resonant levels. The number of resonant levels depends on the height of the barriers. When a bias is applied to the RTD system, as long as the energy of electrons which flow in the RTD is lower than the energy of the resonant level, the most of electrons are reflected by the barrier because of the small transmission coefficient. When the energy of the electrons matches the resonant energy, electrons can go through the barriers efficiently due to resonant tunneling, and the current acquires the maximum value. On the other hand, the current decreases
after the energy of the electrons exceeds the resonant energy.
As a result, RTD systems exhibit characteristic negative differential resistance (NDR) in the current-voltage (I-V) relation
\cite{CEsakiTsuAPL1974}.
Until now, RTDs have been mostly used as high-frequency oscillators device using NDR characteristics
because the tunneling is the fastest charge-transport mechanism in semiconductors
\cite{SollenerAPL1984}.
Resonant tunneling diodes are presently the highest-frequency active semiconductor devices in existence \cite{Ironside2008,Meissner2011,Meissner2012}
and
oscillation frequencies above 1 THz have
recently been realized \cite{AsadaJJAP2008, SuzukiAPL2010, Asada2012}.
%the causes of such phenomena have not yet been definitely identified.

From a fundamental physics point-of-view, the RTD system provides a simple and convenient "context" for studying and testing various methods of analysis for nanoscale quantum devices \cite{Datta,FerryGoodnickBird}.
Frensley discovered NDR in the I-V curve through a numerical computation treating a quantum Liouville
equation in the Wigner representation that adopted open boundary condition and ignored phonon-scattering
processes \cite{FrensleyPRL1986, FrensleyPRB1987, FrensleyRevModPhys1990}. %, Ravaioli1985}.
Kluksdahl et al. incorporated dissipative and self-consistent effects,
employing the Poisson-Boltzmann equation by adopting a relaxation time
approximation, and succeeded in modeling the experimentally observed hysteresis
behavior of the I-V curve \cite{FerryPRB1989}.
Jensen and Buot developed a numerical scheme to treat systems of the same kind
and found evidence that the current oscillation and plateau-like behavior
arise from intrinsic bistability \cite{BuotPRL1991, BuotIEEE1991, Plaummer1996, BuotJAP2000, Yoder2010}.

When plateau-like behavior and hysteresis of the I-V curve in the NDR region, which were thought to arise from the feedback of the electrostatic field, were experimentally observed \cite{GoldmanTsuiPRL1987, Martin1994}, Sollner claimed that they result merely from resonance with the external circuit \cite{SollenerPRL1987}.
Although theoretical calculations have provided evidence of intrinsic bistability and self-excited current oscillations in the NDR region, such phenomena have not been justified by experimental means. In addition, because there exists no well-established methodology that can be applied rigorously to this type of model and includes the effect of dissipation, which is the origin of Joule heat, previous theoretical results have not been well justified. The HEOM approach is ideal to clarify a role of bistability in the NDR region and for detailed analysis of the RTD system.

%In a previous study \cite{STJPSJ}, we demonstrated through consideration of a RTD
%that with the HEOM approach, the effects of non-Markovian thermal fluctuations
%and dissipation at finite temperature on quantum transport can be investigated
% rigorously.
%We found that while most of the current oscillations decay in time in the NDR region,
%there exists persistent oscillation in a plateau of the NDR region.
%In this paper,
%we explore the cause of current oscillation by considering several widths of the
% doped contact regions and
%two different cases for the strength of the system-bath coupling. The value of the
% noise correlation time is also chosen to be twice shorter than the previous study.

This paper is organized as follows.
In Sec. 2, we introduce the reduced
hierarchy equations of motion applicable to the resonant tunneling problem.  
We then present the computational details for the numerical simulations in Sec. 3. 
Numerical results for the I-V curves and current oscillations are presented in Secs. 4.
Section 5 is devoted to concluding remarks. 

%========================================================================
\section{Formulation\label{sec:Formulation}}

We consider the following Caldeira-Leggett Hamiltonian \cite{CLAnnlPhys1983},
which describes the dynamics of an electron subjected to a thermal environment:
\begin{eqnarray}
  \hat{H} = \frac{\hat{p}^2}{2m} + U(\hat{q};t)
   + \sum_j \left[ \frac{\hat{p}_j^{2}}{2m_j} + \frac{m_j\omega_j^2}{2} \left( \hat{x}_j - \frac{a_j V(\hat{q})}{m_j\omega_j} \right)^2 \right]. \label{CLHamiltonian}
\end{eqnarray}
Here, $m$, $\hat{p}$, and $\hat{q}$ are the mass, momentum and position variables of the electron,
and $m_j, \hat{p}_j, \hat{x}_j$ and $\omega_j$ are the mass, momentum, position and frequency variables of the $j$th phonon oscillator mode.
In equation (\ref{CLHamiltonian}), the electron-phonon interaction is given by
\begin{eqnarray}
  \hat{H}_I = - V(\hat{q}) \sum_j a_j \hat{x}_j .
\end{eqnarray}
Here, $V(\hat{q})$ is any function of $\hat{q}$ and the quantities $a_j$ are coefficients that depend on the nature of the electron-phonon coupling.

The heat bath can be characterized by the spectral distribution function, defined by
\begin{eqnarray}
  J(\omega) \equiv \sum_j \frac{a_j^2}{2m_j\omega_j} \delta(\omega-\omega_j),
\end{eqnarray}
and the inverse temperature, $\beta \equiv 1/k_{\mathrm{B}}T$, where $k_\mathrm{B}$ is the Boltzman constant.
We assume the Drude distribution, given by
\begin{eqnarray}
  J(\omega) = \frac{m\zeta}{\pi} \frac{\gamma^2\omega}{\gamma^2 + \omega^2},
\end{eqnarray}
where the constant $\gamma$ represents the width of the spectral distribution of the collective phonon modes
and is the reciprocal of the correlation time of the noise induced by phonons.
The parameter $\zeta$ is related to the electron-phonon coupling strength.
For the collective heat bath coordinate $\hat{X} = \sum_j a_j \hat{x}_j$,
the canonical and symmetrized correlation functions, respectively defined by
$\Psi(t) \equiv \beta \langle \hat{X} ; \hat{X}(t) \rangle_\mathrm{B}$ and
$C(t) \equiv \frac{1}{2} \langle \hat{X}(t) \hat{X}(0) + \hat{X}(0) \hat{X}(t) \rangle_{\mathrm{B}}$, where $\hat{X}(t)$ is the Heisenberg representation of $\hat{X}$,
and $\langle \cdots \rangle_{\mathrm{B}}$ represents the thermal average over the bath modes, are given by \cite{TJPSJ2006,TKJSPS1989}
\begin{eqnarray}
\Psi(t) = { m \zeta \gamma}
 {\rm e}^{ - \gamma \left| t \right|},
\label{eq:L_1barGM}
\end{eqnarray}
and
\begin{eqnarray}
  C(t) = c_0 e ^{ - \gamma \left| t \right|}  + \sum\limits_{k = 1}^\infty  {c_k } e ^{ - \nu _k \left| t \right|}.
\label{eq:L_2GMDef}
\end{eqnarray}
Here, $\nu _k  \equiv 2\pi k/\beta \hbar$ are the Matsubara frequencies, and we have
\begin{eqnarray}
c_0  = \frac
{\hbar m\zeta \gamma ^2 }
{2}
\left[ {\frac{2}
{{\beta \hbar \gamma }} + \sum_{k=1}^\infty
 {\frac{{4\beta \hbar \gamma }}
{{(\beta \hbar \gamma )^2  - (2\pi k)^2 }}} } \right],
\label{eq:c_0}
\end{eqnarray}
and
\begin{eqnarray}
c_k  = - \frac
{\hbar m \zeta \gamma ^2 }
{2}
\frac{{ 8
 \pi k}}
{{(\beta \hbar \gamma )^2  - (2\pi k)^2 }}.
\label{eq:c_k}
\end{eqnarray}
The function $C(t)$ is analogous to the classical correlation function of $X(t)$
and corresponds to the correlation function of the bath-induced noise, whereas $\Psi(t)$ corresponds to dissipation.
The noise $C(t)$ is related to $\Psi(t)$ through the quantum version of the fluctuation-dissipation theorem,
$C[\omega] = \hbar \omega \coth(\beta \hbar \omega/2) /2 \Psi[\omega]$,
which insures that the system exists in the thermal equilibrium state for finite temperatures \cite{KuboToda85}.
Note that in the high temperature limit, $\beta\hbar\gamma \ll 1$,
the noise correlation function reduces to $C(t) \propto {\rm e}^{ - \gamma \left| t \right|}$.
This indicates that the heat bath oscillators interact with the system in the form of Gaussian-Markovian noise.

To derive the equation of motion for the electron,
we use the reduced density operator of the system by taking the trace over the heat bath degrees of freedom:
\begin{eqnarray}
  \hat{\rho}(t) = \mathrm{Tr}_{\mathrm{B}} {\hat{\rho}_\mathrm{tot}(t)}.
\end{eqnarray}
In the path integral representation, the reduced density matrix elements are written
\begin{eqnarray}
 \rho (q,\,q';\,t)= \int {D[q( \tau )]\int {D[q'( \tau )]\int {dq_i } \int {dq'_i } } } \rho (q_i ,q'_i )\rho_{\mathrm{CS}} (q,\,q_i,\,q',\,q'_i;\,t) \nonumber \\ 
   \times {\rm e} 
^{\frac{\rm i}
{\hbar }S_A [q;\,t]}  
F[q,\,q';\,t]{\rm e}^{ - \frac{\rm i}
{\hbar }S_A [q';\,t]},
\label{eq:rho_reduced}
\end{eqnarray}
where $S_A [q;\,t] $ is the action for the Hamiltonian of the system, $\hat{H}_A=p^2/2m+U(q; t)$, expressed as
\begin{eqnarray}
S_A [q;\,t] \equiv  \int_{t_i}^{t} d\tau \left[ \frac{1}{2} m \dot q^2(\tau)   - U(q (\tau); \tau) \right],
\label{eq:actionfunc}
\end{eqnarray}
$\rho (q_i ,q'_i )$ is the initial state of the system at time $t_i$,
$F[q,\,q';\,t]$ is the influence functional \cite{Feynman63},
and $\rho _{\mathrm{CS}}(q,\,q_i,\,q',\,q'_i;\,t)$ is the initial correlation function between 
the system and the heat bath \cite{Grabert88}.
The functional integrals for $q(\tau)$ and $q'(\tau)$
are carried out from $q(t_i)=q_i$ to $q(t)=q$ and from $q'(t_i)=q_i'$ to $q'(t)=q'$, respectively.
In the HEOM approach, we can specify $\rho_{\mathrm{CS}} (q,\,q_i,\,q',\,q'_i;\,t)$ by nonzero hierarchy elements. 
To simplify the derivation of the HEOM, here we set $\rho_{\mathrm{CS}} (q,\,q_i,\,q',\,q'_i;\,t)=1$ and regard $\rho(q_i ,\,q'_i)$ as a temporal initial condition. Then, after deriving the HEOM, we take into account $\rho_{\mathrm{CS}} (q,\,q_i,\,q',\,q'_i;\,t)$ through implementation of a hierarchy of initial conditions that can be evaluated numerically \cite{TJPSJ2006}. The influence functional for the inverse temperature $\beta$ is given by \cite{Feynman63,Grabert88}
\begin{eqnarray}
F[q,\,q';\,t] ={\rm exp}\left\{ \left( { - \frac{\rm i}{\hbar }} \right)^2 \int_{t_i }^t {d\tau } V^{\times}(q,\,q';\,\tau) \right. \nonumber \\
\times \left. \left[ \frac{\partial }{\partial \tau }\int_{t_i }^{\tau } {d\tau' } \,\frac{i \hbar}{2} \Psi (\tau - \tau' )
V^{\circ}(q,\,q';\,\tau') 
  + \int_{t_i }^{\tau } {d\tau' } C(\tau  - \tau' )
V^{\times}(q,\,q';\,\tau')  \right] \right\},  \nonumber \\
\label{eq:influenceq}
\end{eqnarray}
where $V^{\times}(q,\,q';\,\tau) \equiv {V\left( {q(\tau )} \right) - V\left( {q'(\tau )} \right)}$
and $V^{\circ}(q,\,q';\,\tau) \equiv {V\left( {q(\tau )} \right) + V\left( {q'(\tau)} \right)}$.
If we choose $K$ so as to satisfy $\nu_K=2\pi K/(\beta\hbar) \gg \omega_c$, where $\omega_c$ is
the characteristic frequency of the system such as the frequency of self-excited current oscillations,
the factor ${\rm e}^{-\nu_k |t|}$ in equation (\ref{eq:L_2GMDef}) can be replaced with Dirac's delta function,
using the approximation $\nu_k{\rm e}^{-\nu_k |t|}\simeq \delta( t ) \quad ({\rm for} \ \ k \geq  K+1)$.
Therefore, $C(t)$ can be expressed as
\begin{eqnarray}
C(t) = c_0 e ^{ - \gamma |t|}  + \sum\limits_{k = 1}^{K}  {c_k } e ^{ - \nu _k |t|} 
+  \delta(t) \sum\limits_{k = K+1}^{\infty} \frac{c_k}{\nu_k} .
\label{eq:L_2GMDef2}
\end{eqnarray}
By choosing $2\pi K \gg \beta\hbar \omega_c$, the above expression allows us to evaluate $C(t)$ for finite $K$ with negligible error at the desired temperature $1/\beta$.

The reduced hierarchy equations of motion (HEOM) can be obtained by considering the time derivative of the reduced density matrix with the kernel given in equations (\ref{eq:L_1barGM}) and (\ref{eq:L_2GMDef2}).
The HEOM have been used to study chemical reactions \cite{Tanimura91, Tanimura92, Ishi05,shi09},
linear and nonlinear spectroscopy \cite{ ITJSPS2005,TM97,Tanimura00,Kato02,Kato04, Ishi07,STJPCA2011,Yan09,Yan10,Yan12},
exciton transfer \cite{IshiFlem09,shi11,Kramer12,Schulten12,DijkstraNJP10,DijkstraNJP12},
electron transfer \cite{Xu07,Tanaka2009JPSJ, Tanaka2010JCP,TJCP2012},
quantum dots \cite{YanQdot08,YanQdot12}, quantum ratchet\cite{KTJPCB2013},
and quantum information \cite{Dijkstra10,Dijkstra12,Nori12}.
A variety of numerical techniques have been developed for the HEOM approach in order to accelerate numerical calculations \cite{Shi09,YanPade10A,YanPade10B,GPU11,Aspru11,Schuten12, Cao2013, Shi2013}.
The accuracy of the HEOM approach has been justified for a Brownian oscillator system \cite{Tanimura00, Kato02,Kato04,Ishi07, STJPCA2011}, a displaced Brownian oscillators system \cite{TM97}, and a spin-Boson system \cite{ITJSPS2005, Yan09, Dijkstra12} via linear and nonlinear response functions by comparing the analytical solutions of the response functions \cite{TM2D1993,TMPRE1993,TOJCP1997}. The validity of the HEOM are also confirmed with other numerically methods such as the iterative quasi-adiabatic propagator path-integral scheme and a time-convolution less master equation in the relevant crossover regime from weak to strong system-bath coupling \cite{Makri2010, Thrwart2011, ITCP2008}.

The HEOM are ideal for studying quantum transport systems,
in conjunction with the Wigner representation,
characterized by the Wigner distribution function,
\begin{eqnarray}
W(p,q;t) \equiv \int_{-\infty}^\infty dr e^{-\frac {i pr}{\hbar}} \rho\left(q+\frac{r}{2},q-\frac{r}{2};t \right),
\end{eqnarray}
because they allow us to treat continuous systems utilizing open boundary conditions and periodic boundary conditions \cite{FrensleyRevModPhys1990,Ravaioli1985}.
Although the Wigner distribution function is not positive definite, it is the quantum analogue of the classical distribution function in the phase space \cite{FrensleyPRL1986, FrensleyPRB1987, FrensleyRevModPhys1990, FerryPRB1989, BuotPRL1991, BuotIEEE1991, Plaummer1996, BuotJAP2000, Cao2004, Yoder2010, Ravaioli1985, Jacoboni2004, JiangCaiTsu2011}. 
Its classical limit can be computed readily.
This is helpful, because knowing the classical limit allows us to identify the purely quantum mechanical effects \cite{Tanimura91,Tanimura92,STJPCA2011,KTJPCB2013, Cao2004}.

While we can handle any form of $V(q)$ \cite{Tanimura00,Kato02,Kato04,Ishi07,STJPCA2011}, here we consider the {linear-linear} system bath coupling case defined by $V(q)=q$. In the Wigner representation, the equations of motion are expressed in hierarchical form as follows \cite{TJPSJ2006,STJPCA2011}: 
\begin{eqnarray}
  \frac{\partial}{\partial t} W^{(n)}_{j_1, \dots, j_K}(t)
  = - \left[ \hat{L}_\mathrm{qm} + \hat{\Xi}' + n\gamma + \sum_{k=1}^K j_k \nu_k \right]W^{(n)}_{j_1, \dots, j_K}(t) \nonumber \\
   + \hat{\Phi}\left[W^{(n+1)}_{j_1, \dots, j_K}(t) + \sum_{k=1}^K W^{(n)}_{j_1, \dots, (j_k+1), \dots j_K}(t) \right] \nonumber \\
   + n\gamma\hat{\Theta}_0 W^{(n-1)}_{j_1, \dots, j_K}(t) + \sum_{k=1}^K j_k \nu_k \hat{\Theta}_k W^{(n)}_{j_1, \dots, (j_k+1), \dots, j_K}(t) \label{HEOM}
\end{eqnarray}
for non-negative integers $n,j_1,\dots, j_K$, where we have chosen $K$ such that $\nu_{{K}} \gg \omega_c$.
In equation (\ref{HEOM}), $-\hat{L}_{\mathrm{qm}}$ is the quantum Liouvillian in the Wigner representation, given by
\begin{eqnarray}
  -\hat{L}_{\mathrm{qm}} W(p,q) \equiv -\frac{p}{m} \frac{\partial}{\partial q}W(p,q)
   - \frac{1}{\hbar}\int_{-\infty}^\infty \frac{dp'}{2\pi\hbar} U_{\mathrm{W}}(p-p',q; t) W(p',q),
\label{Liouville1}
\end{eqnarray}
with
\begin{eqnarray}
  U_{\mathrm{W}}(p,q;t) \equiv 2\int_0^\infty dr \sin\left( \frac{pr}{\hbar} \right) \left[ U\left(q+\frac{r}{2}; t\right) - U\left(q-\frac{r}{2}; t \right) \right]. \label{Wigner_pot}
\label{Liouville2}
\end{eqnarray}
The other operators appearing in equation (\ref{HEOM}) are the bath-induced relaxation operators, defined as
\begin{eqnarray}
  \hat{\Phi} \equiv \frac{\partial}{\partial p},
\end{eqnarray}
\begin{eqnarray}
  \hat{\Theta}_0 \equiv \zeta \left[p + \frac{m\hbar\gamma}{2} \cot \left( \frac{\beta\hbar\gamma}{2}\right) \frac{\partial}{\partial p}\right],
\label{HEOMT1}
\end{eqnarray}
\begin{eqnarray}
  \hat{\Theta}_k \equiv \frac{c_k}{\nu_k} \frac{\partial}{\partial p},
\label{HEOMT2}
\end{eqnarray}
and
\begin{eqnarray}
  \hat{\Xi}' \equiv \left\{-\frac{m\zeta}{\beta} \left[1 - \frac{\beta\hbar\gamma}{2}\cot\left( \frac{\beta\hbar\gamma}{2} \right) \right]
   + \sum_{k=1}^K  \frac{c_k}{\nu_k}  \right\} 
%#################################################3
\frac{\partial^2 }{\partial p^2}.
%#############################################
\label{HEOMT3}
\end{eqnarray}
In the case that the quantity $N \equiv n + \sum_{k=1}^K j_k$ satisfies the relation
$N \gg {\omega_c}/{\min(\gamma,1/\beta\hbar)}$,
this infinite hierarchy can be truncated with negligible error at the desired temperature $1/\beta$ by the terminator \cite{STJPCA2011}
\begin{eqnarray}
  \frac{\partial}{\partial t} W^{(n)}_{j_1, \dots, j_K}(t) = - (\hat{L}_{\mathrm{qm}} + \hat{\Xi}') W^{(n)}_{j_1, \dots, j_K}(t).
\end{eqnarray}
The validity of the above truncation scheme and its extension for efficient numerical calculations have been discussed for the spin-Boson system \cite{Shi09,YanPade10A,YanPade10B}.
Note that only $W^{(0)}_{0,\dots,0}(p,q;t) \equiv W(p,q;t)$ has physical meaning,
and the other elements $W^{(n)}_{j_1, \dots, j_K}(p,q;t)$ with $(n;j_1, \dots, j_{{K}}) \ne (0;0,\dots,0)$ are auxiliary operators
introduced to avoid the explicit treatment of the inherent memory effects that arise in the time evolution of the reduced density matrix.
If the noise correlation is very short ($\gamma \rightarrow \infty$) and the temperature is high
(i.e., the noise is white),
the quantum Fokker-Planck equation can be derived in a form similar to that of Kramers equation \cite{CLPhysica1983,Waxman1985,Cao1997}.
In the present case, however, we cannot employ the white noise approximation,
because quantum effects play a dominant role in the low temperature regime ($\beta\hbar\omega_c \gg 1$) \cite{ITJSPS2005, KTJPCB2013}.
%Although the original quantum Fokker-Planck equation, derived by Caldeira 
%and Leggett,is valid only for sufficiently 
%high temperatures ($\beta\hbar\omega_c \ll 1$)  \cite{CLPhysica1983},
%our formulation is valid for arbitrary temperature {$1/\beta$ 
%with negligible error.}

In equation (\ref{Wigner_pot}), $U(\hat{q};t)$ is the effective potential for the electron,
which can be written \cite{FerryPRB1989, BuotPRL1991, BuotIEEE1991, Plaummer1996, BuotJAP2000, Yoder2010}
\begin{eqnarray}
 U(\hat{q};t) = U_{\mathrm{static}}(\hat{q}) + U_{\mathrm{self}}(\hat{q};t),
\label{effectivepote}
\end{eqnarray}
where $U_{\mathrm{static}}(\hat{q})$ and $U_{\mathrm{self}}(\hat{q};t)$ are the static and self-consistent parts, respectively. 
As the static potential, we employ the double-barrier structure depicted in figure 1 (b).
The self-consistent part, $U_{\mathrm{self}}(q;t) = - e\phi(q;t)$, is calculated
from the electron distribution at each time step in the integration of equation (\ref{HEOM}) using the Poisson equation
\begin{eqnarray}
  - \frac{\partial}{\partial q} \left[\epsilon \phi(q;t) \right] = e \left[n^{+}(q) - P(q,t) \right],
\label{Poisson}
\end{eqnarray}
where $\epsilon$ is the dielectric constant, 
$n^{+}(q)$ is the doping density
and $P(q;t) = \int_{-\infty}^{\infty}dpW^{(0)}_{0, \dots,0}(p,q;t) /(2\pi\hbar) $ is the electron density
calculated from the Wigner distribution.
Coupling the HEOM to the Poisson equation,
we obtain a fully self-consistent model of quantum electron transport.
This allows us to examine charge redistribution effects.

%========================================================================
\section{Computational details\label{sec:detail}}

The equations of motion given in (\ref{HEOM}) were numerically evaluated using finite mesh representations
of the Wigner distribution functions. The spatial derivative of the kinetic term in the Liouville operator, $-(p/m) \partial W(p,q)/\partial q$,
was approximated by using a third-order left-handed or right-handed difference scheme.  Note that the first-order difference scheme is not sufficiently accurate for the present problem, because this scheme introduces
% artificial viscosity
{false diffusion} for the wavepacket dynamics, which {suppresses} the self-excited current oscillations.
Depending on the sign of the momentum, the expressions are given by
\begin{eqnarray}
  \frac{\partial W(p_k,q_j)}{\partial q} 
  =\frac{1}{6\Delta q} \left( {2}W(p_k,q_{j+1}) \right. &+& 3W(p_k,q_j)  \nonumber \\
&-& \left. 6W(p_k,q_{j-1}) + W(p_k,q_{j-2})  \right)
\label{left-handed}
\end{eqnarray}
for $p_k > 0$, and  
\begin{eqnarray}
   \frac{\partial W(p_k,q_j)}{\partial q} =
   \frac{1}{6\Delta q} \left(-\right. &W&(p_k,q_{j+2}) + 6W(p_k,q_{j+1}) \nonumber \\ 
&-& \left. 3W(p_k,q_j) {-} 2W(p_k,q_{j-1}) \right)
\label{right-handed}
\end{eqnarray}
for $p_k < 0$, in order to treat continuous systems utilizing the inflow and outflow boundary conditions with use of the first-order difference scheme at $q=L$ ($p > 0$) and $q=0$ ($p < 0$)\cite{FrensleyPRL1986, FrensleyPRB1987, FrensleyRevModPhys1990}.

The inflow boundary conditions were set by stipulating that
$W^{(n)}_{j_1, \cdots , j_{{K}}} (p<0, q =L)$ and
$W^{(n)}_{j_1, \cdots , j_{{K}}} (p>0, q=0)$
are given by the equilibrium distribution of a free particle
calculated from the HEOM with periodic boundary conditions. 
Due to fluctuations and dissipation, the flow of a wavepacket reaches a steady state
even when there exists a non-zero bias voltage.
The validity of the boundary conditions was verified by considering several system sizes. 

Other derivatives with respect to $p$ were approximated using the fourth-order centered difference scheme given by
\begin{eqnarray}
\frac{\partial W(p_k,q_j)}{\partial p}
  = \frac{1}{12\Delta p}\left( - \right. &W&(p_{k+2},q_j) + 8W(p_{k+1},q_j) \nonumber \\  &-& \left. 8W(p_{k-1},q_j) + W(p_{k-2},q_j) \right) ,
\end{eqnarray}
and
\begin{eqnarray}
  \frac{\partial^2 W(p_k,q_j)}{\partial p^2}
  = \frac{1}{12\Delta p^2} \left(-\right. &W&(p_{k+2},q_j) + 16W(p_{k+1},q_j)- 30W(p_k,q_j)  \nonumber \\
&+& \left. 16W(p_{k-1},q_j) - W(p_{k-2},q_j) \right) .
\end{eqnarray}
The mesh size for the position, $\Delta q$, and momentum,$\Delta p/(2\pi\hbar)$, are respectively 0.2825 nm and 3.540 $\mathrm{nm}^{-1}$.

We set the parameters used in the HEOM as $\gamma = 24.2$ THz ($\gamma^{-1} = 4.13$ fs),
$\zeta = 72.5$ GHz ($\zeta^{-1} = 13.8$ ps), and $T = 300$ K
in order to create conditions close to those used in previous theoretical studies.
In Appendix B, we also report the results of calculations for the strong coupling case,
with $\zeta=120.8$ GHz ($\zeta^{-1}=8.28$ ps),
to elucidate the role of dissipation.
The depth of the hierarchy and the number of Matsubara frequencies were chosen as $N \in \{2-6\}$ and $K \in \{1-3\}$, respectively.
To model GaAs, the effective mass of the electron was assumed to be constant across the device and equal to $0.067 m_0$, where $m_0$ is the electron mass in vacuum.
The dielectric constant in equation (\ref{Poisson}) was set as $\epsilon=12.85$.

As the static double-barrier potential,
which models the hetero-structure of GaAs sandwiched between two thin AlGaAs layers,
we set the widths of quantum well (undoped GaAs), barrier (undoped AlGaAs), and spacer layer (undoped GaAs)
to be 4.520 nm, 2.825 nm, and 2.825 nm, respectively.
The height of the potential barriers was 0.27 eV.
The conduction band edge consists of a single quantum well bounded by tunneling barriers (figure 1(b)).
The widths of the contact regions (the yellow parts in figure 1(a),
where GaAs is doped with a concentration of $2\times10^{18}$ $\mathrm{cm}^{-3}$)
were chosen as 16.950 nm, 33.900 nm, and 42.375 nm. 
{Note that, to adapt the one dimensional model,
we rescaled the concentration by multiplying the doping density by unit area.}

In a previous study \cite{STJPSJ}, we chose a smaller value of $\gamma$, $\gamma = 12.1$ THz ($\gamma^{-1} = 8.26$ fs),
and fixed the width of the contact region as 42.375 nm. {Note that, since the effective system-bath coupling strength is estimated as $\propto {\zeta}{\gamma^2\omega_c}/(\gamma^2 + \omega_c^2)$, where $\omega_c$ is the characteristic frequency of the system \cite{Tanimura92}, the damping strength in the present case is slightly larger than the previous case.}
In that case, we found hysteresis, double plateau-like behavior, and self-excited current oscillation
in the negative differential resistance (NDR) region of the current-voltage curve.
We found that while most of the current oscillations decay in time in the NDR region,
there exists a non-transient oscillation characterized by a tornado-like rotation in the Wigner space
in the upper plateau of the NDR region.
In this paper, we explore the cause of such current oscillations
by considering several values of the width of the contact regions
in cases of both weak and strong coupling, characterized by different values of $\zeta$.

%========================================================================

\section{Results\label{sec:Results}}
We determined the characteristics of the current-voltage (I-V) according to the following procedure.
First, we integrated equation (\ref{HEOM}) at zero bias voltage without the self-consistent part of the effective potential under the inflow boundary conditions specified above.
When we obtained the \textit{temporal} steady state,
the obtained distribution was then used as the initial distribution for the self-consistent calculation,
and we then integrated equation (\ref{HEOM}) again,
with the effective potential $U(\hat{q};t)$ evaluated iteratively using the Poisson equation given in (\ref{Poisson}).
Under this procedure, when the distributions reached the \textit{genuine} steady state
$W^{(n)}_{j_1, \dots, j_K}(p,q; t\rightarrow \infty)$, the current was calculated 
by $I(t)=\int dp\ pW_{0,\dots,0}^{(0)}(p,q;t)/2\pi\hbar m$
%using the 
%distribution $W_{0,\dots,0}^{(0)}(p,q; t\rightarrow \infty)$,
and then the \textit{genuine} state was used as the initial distributions for the next bias step.
{
While the  \textit{temporal} steady states were obtained for the static potential, $U_{\mathrm{static}}(\hat{q})$, the  \textit{genuine} steady states were calculated from the effective potential, $U_{\mathrm{static}}(\hat{q}) + U_{\mathrm{self}}(\hat{q};t)$. Since the value of the effective potential depended on the hysteresis of a physical process and since we wanted to use a uniquely determined steady state, we chose the \textit{temporal} steady state as a temporal initial state to have the  \textit{genuine} steady state.}

Following the above steps, we increased the bias from 0.000V to 0.500V,
and then decreased it to 0.000V with bias steps of 0.01 V in the normal region
and 0.002 V in the negative differential resistance (NDR) region.
{The corresponding sweeping rates were
$5 \times 10^9 \ \mathrm{V/s}$ in the normal region
and $5 \times 10^7 \ \mathrm{V/s}$ in the NDR region.}
{We found that the profiles of the I-V curves did not change
for slower sweeping rate than the present values, whereas
the width of plateau observed in the NDR region often became smaller for a faster sweeping rate.} At each step, we integrated the equation of motion until the system 
exhibited the steady current.
%reached the steady state distribution.
However, in some cases, in the NDR region, steady current oscillation arose.
%, and a steady state was thus not realized.
In such cases, the value of current in figure 2(a) was evaluated 
as a time average after stable oscillations were realized
(between 30 to 40 ps in most cases).
%as a time average between 30 ps to 40 ps.

\begin{figure}
\begin{center}
  \includegraphics[width=100mm]{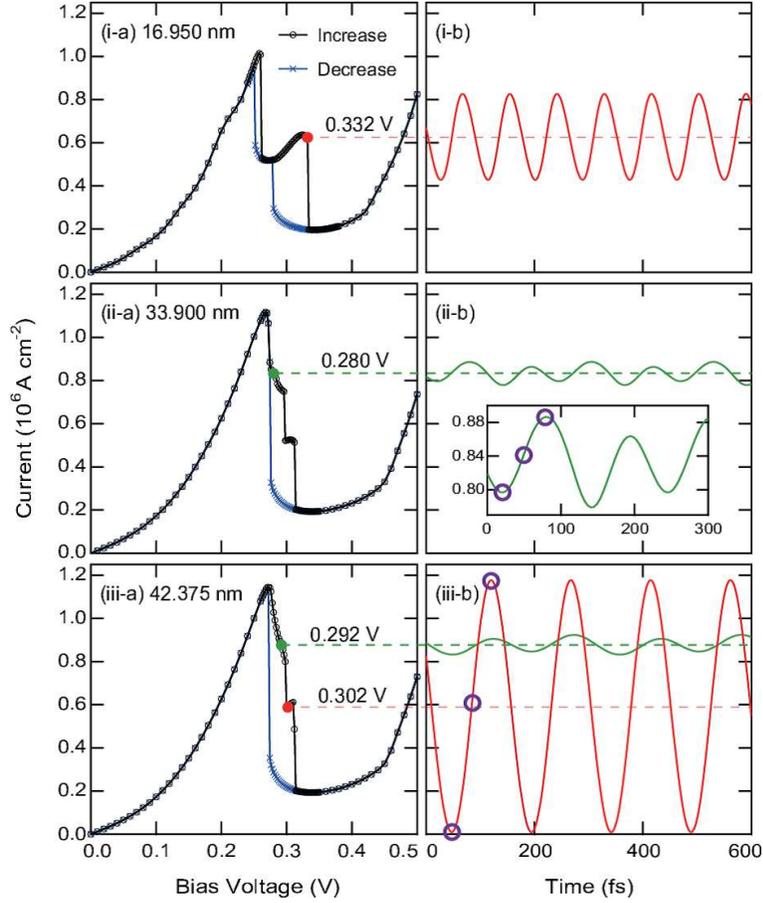}
\end{center}
\caption{\label{fig2}
(a) The I-V characteristics for three sizes of the contact regions: (i) 16.950 nm; (ii) 33.900 nm; (iii) 42.375 nm.
The black curve with circles represents the case in which the bias is increasing,
and the blue curve with the $\times$s represents the case in which the bias is decreasing.
(b) Time evolution of the current for the value of the bias voltage at which current oscillation occurs.
The red curves represent the current oscillation with large amplitude,
and the green curves represent the current oscillation with small amplitude.
The inset in (ii-b) contains a close-up view of one portion of the current. 
The snapshots of the Wigner distribution at the time points marked with purple circles
are presented in figures 4 and 5.}
\end{figure}

In figure 2, we present (a) current-voltage (I-V) relations and
(b) the time evolution of the self-excited current oscillation in the weak coupling case ($\zeta = 72.5$ GHz)
for three values of the width of the contact regions (the yellow parts in figure 1(a)).
In these plots, the sizes of the contact regions are (i) 16.950 nm, (ii) 33.900 nm, and (iii) 42.375 nm, respectively,
with a fixed doping concentration of GaAs.
These graphs reveal NDR behavior, hysteresis, and plateaus in the I-V curve.
Moreover, self-excited current oscillation appears in some regions of the plateau.
While Jensen and Buot observed only a single plateau similar to that in figure 2(i-a)
with current oscillation \cite{BuotPRL1991},
our results in figure 2(ii-a) and 2(iii-a) exhibit a double plateau structure.
While the experimental result shown by \cite{GoldmanTsuiPRL1987} is similar to figure 2(i-a), those shown by \cite{Ironside2008, AsadaJJAP2008, SuzukiAPL2010} are similar to figures 2(ii-a) and 2(iii-a).
For the sake of comparison, we present a graph corresponding to figure 2 calculated from the Boltzmann equation in Appendix C.
In the case of a single plateau structure ({Figure 2(i-a)}),
the width of the plateau is large
% for a narrow contact region,
, while in the case of a double plateau structure ({Figures 2(ii-a) and (iii-a)}),
the width is small
% for a wide contact region. 
. The I-V curves obtained in the strong coupling case ($\zeta=120.8$ GHz) are presented in Appendix B.
In that case, NDR behavior, hysteresis, and a single plateau in the I-V curve are observed,
but steady current oscillation does not appear, even in the plateau.

\begin{figure}
\begin{center}
  \includegraphics[width=120mm]{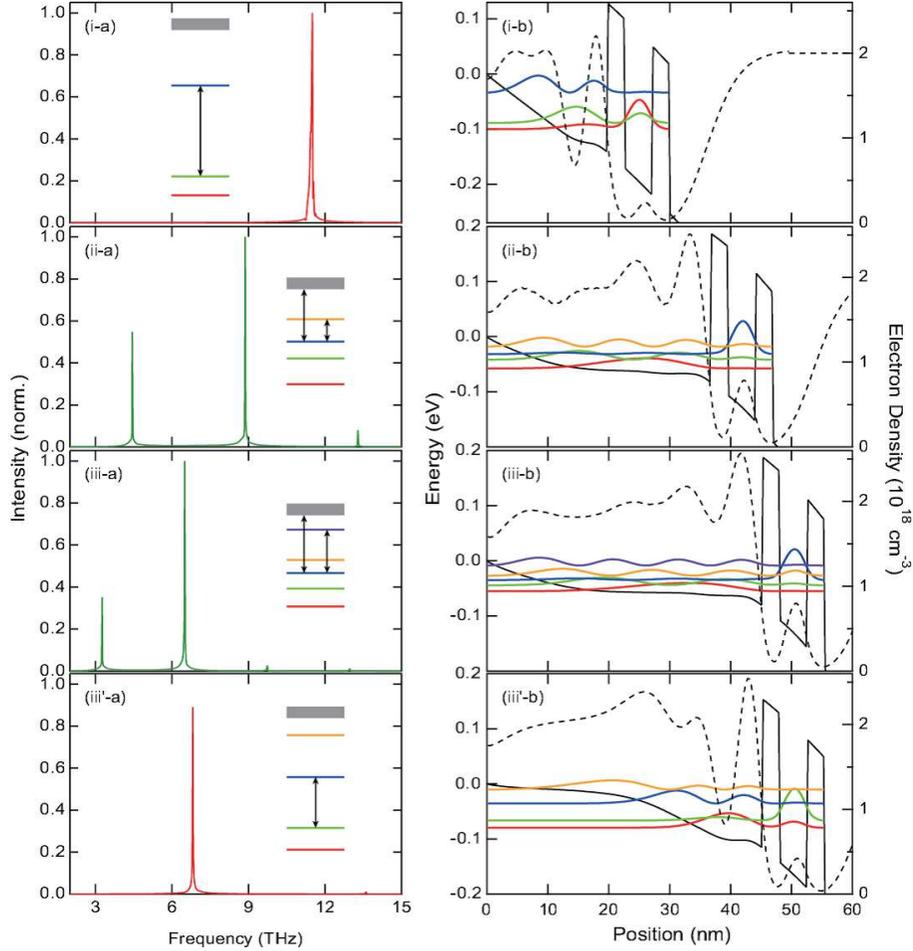}
\end{center}
\caption{\label{fig3}
(a) The frequency distribution of the current oscillation that arises in the plateau region in the case of increasing bias.
The size of contact region and the bias voltage are as follows:
(i) 16.950 nm and 0.332 V (red curve in figure 2(i-b));
(ii) 33.900 nm and 0.280 V (green curve in figure  2(ii-b));
(iii) 42.375 nm and 0.292 V (green curve in figure  2(iii-b));
(iii$'$) 42.375 nm and 0.302 V (red curve in figure  2(iii-b)).
The insets depict the corresponding transitions.
The colored lines represent the eigenstates in the emitter basin given in the right graph,
while the thick grey line represents the continuous energy band.
(b) The time-averaged effective potential (black solid curve) and time averaged electron density (black dashed curve) for (i)-(iii').
The red, green, blue, orange, and purple curves represent the eigenstates in order of increasing eigenenergy calculated using the averaged effective potential without the heat bath.
{Basin-like structures denoted in the red squares are formed on the emitter side of the potential (emitter basin).}
}
\end{figure}

As in the case considered in the previous paper \cite{STJPSJ},
we find that most of the current oscillations decay in time in the NDR region,
but there also exist non-decaying oscillations in some regions of the plateau,
as seen in figures 2 (i-b) - (iii-b).
The Fourier components of each persistent oscillation are plotted in figures 3 (i-a) - (iii'-a).
We can classify these oscillations into two types, according to the current amplitude.
The first type is observed in the single plateau (figure 2(i))
and the lower part of the double-plateau (figure 2(iii)) with large amplitude (red).
The plateau in this case is located in the middle of the NDR region.
The second type is observed in the upper part of the double-plateau (figures 2(ii) and 2(iii))
with small amplitude (green).
The plateau in this case is located at a current approximately three-quarters of the peak current.
The second type of oscillation contains two Fourier components (figures 3 (ii-a) and (iii-a)),
whereas the first type contains just a single component (figures 3 (i-a) and (iii'-a)).
We find that as the size of the contact regions increases, the frequency of each peak decreases.

As mentioned by Kluksdahl \textit{et al.} \cite{FerryPRB1989} and Zhao \textit{et al.} \cite{BuotJAP2000},
the quantum well formed on the emitter side of the effective potential plays an important role in the realization of hysteresis and plateau-like structure in the NDR region.
The time averaged electron densities (black dashed curves) and the effective potentials (black solid curves)
calculated from the Wigner distribution function in the case of increasing bias are plotted in figures 3 (i-b) - (iii'-b).
For reference, in Appendix A,
we present a graph corresponding to figure 3(iii'-b) depicting the situation in the case of decreasing bias.
A basin-like potential on the emitter side (emitter basin) is observed in the case of increasing bias,
while the emitter basin does not exist in the case of decreasing bias.
In the case of increasing bias, when the bias exceeds the peak point of the I-V curve,
the kinetic energy of the inflowing electron becomes larger than the eigenenergy of the resonant tunneling state.
As a result, the reflection of the current from the emitter side of the barrier becomes large,
and the electron distribution function becomes concentrated near the barrier, as depicted in figure 3 (i-b) - (iii'-b).
When the electron density increases, the effective potential decreases.
As a result, the emitter basin appears.
When the emitter basin becomes sufficiently {deeper},
there appear resonant tunneling states between the emitter basin and the double-barrier well.
In such a case, we find current oscillation and a plateau of the I-V curve in the NDR region.
In the case of decreasing bias, however, because there is no resonant tunneling state between the emitter basin and double-barrier well,
the current is much smaller than in the case of the increasing bias.
This difference causes the hysteresis behavior.

To elucidate this point more clearly, we solve the {steady-state} Schr\"{o}dinger equation
for the {regions of} emitter basin and the double-barrier well,
to obtain approximate eigenstates and eigenenergies of an electron whose energy is lower than the continuous energy band on the emitter side.
{Since bound states are not formed in the collector side of the potential, we exclude this region from the calculations.}
It should be noted that we employ the time-averaged potential for the purpose of the graph,
but the effective potential and the corresponding eigenstates actually vary in time,
because they depend on the electron distribution function, and it varies in time.
Thus, for example, the identifications of the first (red) and second (green) eigenstates and the third (blue) and fourth (orange) eigenstates change in time, often becoming degenerate and interchanging.
In addition, we ignore the continuous band on the collector side
and the influence of the heat bath when calculating the eigenstates and eigenenergies.
For this reason, the calculated eigenenergies are not precise,
but we find that they are sufficient for determining the cause of the current oscillation,
because each resonant frequency is rather isolated when estimating the oscillation frequency.

We find that each oscillation peak in both the large and small oscillation cases can be attributed
to transitions between eigenstates in the emitter basin,
as depicted in the insets of figures 3(i-a)-3(iii'-a).
As shown in figures 3 (i-b) and 3 (iii'-b), the {first and second eigenstates are} the tunneling states in the large oscillation case,
while the third eigenstate in figures 3 (ii-b) and (iii-b) is the tunneling state in the small oscillation case.
When we compare the profiles of each eigenstate and the electron density distributions,
we find that the tunneling state and the higher energy eigenstate close to the tunneling state are populated in both cases.
Thus, we conclude that the current oscillation results from transitions between these two states,
with the frequency of the oscillation determined by the frequency of these transitions.
{Since both the first and second eigenstates change in time, the amplitude of current becomes large in the large oscillation case.}
%Because 
When
the structure of the emitter basin becomes
%is 
stable with respect to change of the bias voltage, a plateau forms.
As the size of the contact regions increases, the transition frequencies decrease,
because the size of the emitter basin increases, while the depth of the basin does not change.
This accounts for the peak shift seen in figures 3 (i-a) and 3 (iii'-a) and figures 3 (ii-a) and 3 (iii-a).
For a small size of contact regions, 
%Because 
the emitter basin becomes more stable with respect to change in the 
bias voltage 
%as the contact regions becomes smaller,
and as the result the plateau becomes larger. 
%when we decrease the size of contact regions.

Both figures 2 (ii) and (iii) exhibit double plateau-like features in the NDR region,
but we find current oscillation only in the upper plateau in the case of figure 2 (ii).
This is because the dissipation in the large oscillation case of figure 2(ii) is not sufficiently strong
to create significant population of the tunneling states in the case of a small basin,
for which the resonant frequencies between the tunneling states and adjacent states are large.
In our previous study \cite{STJPSJ}, we considered the same condition as in figure 2 (iii),
with a value of $\gamma$ half as large ($=12.1$ THz).
In that case, however, there was only upper plateau oscillation,{ while we observed the bistability and the double plateau-like feature.}
This is because the effective system-bath coupling strength
{in the previous case is weaker than in the present case \cite{Tanimura92}.}
Thus, in that case, the ground tunneling states are not populated from the conduction band through dissipation.
Note, however, that if the system-bath coupling is too strong, dissipation suppresses the current oscillation,
as explained in Appendix B.

\begin{figure}
\begin{center}
  \includegraphics[width=70mm]{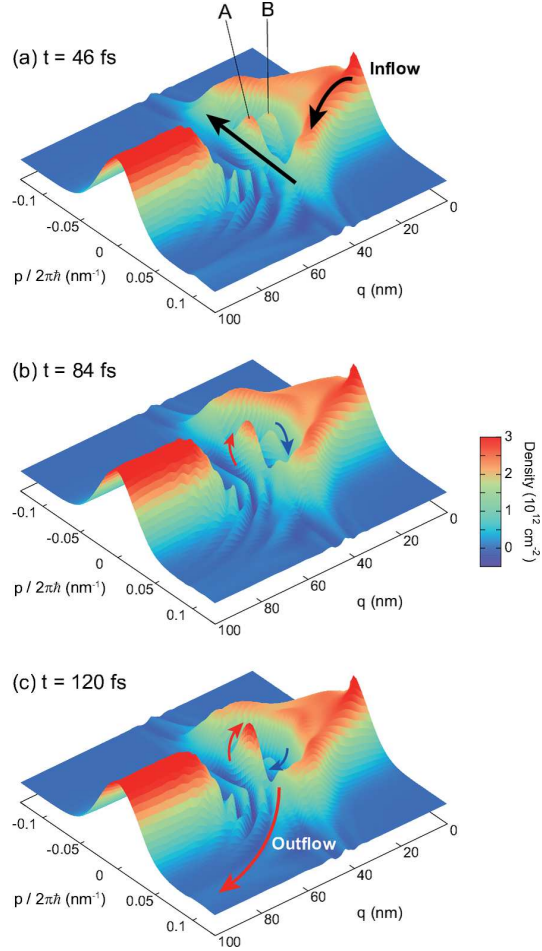}
\end{center}
\caption{\label{fig4}
Current oscillation with a large amplitude observed in the lower part of the double-plateau structure
at the times marked on the red curve in figure  2 (iii-b) depicted as snapshots of the Wigner distribution.
(a) Current flows into the system from the emitter side of the boundary,
and then it is scattered ($p_{in} \rightarrow -p_{in}$, where $p_{in}$ is the momentum of the inflow current) by the emitter side of the barrier almost elastically.
The scattered electron flows into peak B.
(b) Peaks A and B become higher and lower in turn, exhibiting motion similar to that of pistons in a two-piston engine.
(c) The current becomes large whenever peak A becomes large due to the tunneling.
}
\end{figure}

One important aspect of the present methodology is that it allows elucidation of the dynamical behavior of the system
through the time evolution of the Wigner distribution function.
We have been able to characterize the patterns of the time evolution of the Wigner distribution for two types of current oscillations, with large and small oscillation amplitude.
Here, we describe these two types in detail, using as one reference,
the large oscillation case in figure 2 (iii-a)
and the small oscillation case in figure 2 (ii-a).
In figure 4, we display snapshots of the Wigner distribution function for the case of large oscillation
at the times marked on the red curve in figure  2 (iii-b).
As illustrated in figures 3 (i-b) and 3 (iii'-b), the characteristic feature of this type
of oscillation is the large electron density near the emitter side of the barrier,
which is observed as a distinct peak separated from the conduction band (at $q= $ 18 nm in figure 3 (i-b) and at $q= $ 43 nm in figure 3 (iii'-b)).
As a result of this feature, the effective potential possesses a deep emitter basin next to the barrier.
The profiles of the eigenstates depicted in figure 3 (iii'-b) indicate that this peak consists of the first (red) to fourth (orange) eigenstates.
We find that a small peak near the edge of the conduction state (at $q= $ 35 nm in figure 3(iii'-b)),
which arises from the third (blue) and fourth (orange) eigenstates in figure 3(iii'-b),
also plays an important role in the current oscillation.
In the Wigner distribution plotted in figure 4(a), these two peaks are denoted by A and B, respectively.
In the situation depicted in figure 4 (a), the current flows into the system from the emitter side of the boundary,
and then it is scattered by the emitter side of the barrier almost elastically,
because the kinetic energy of the current electron is much higher than that of the tunneling state.
The scattered current is trapped by the emitter basin and rotates clockwise around the peaks A and B,
but, due to dissipation, some of it flows into peak B while losing energy.
In the eigenstate representation given in figure 3(iii'-b),
this behavior corresponds to population transfer from the fourth (orange) to the third (blue) eigenstate.
In figure  4 (b), when the third (blue) state decays further to the first (red) and second (green) tunneling states,
the height of A increases.
As a result, peak A becomes higher, while peak B becomes lower.
In figure 4(c), because peak A is related to the tunneling state,
the outflow current becomes larger whenever peak A becomes higher.
Throughout this process, the peaks A and B become higher and lower by turns,
in a manner reminiscent of the piston in a two-piston engine.
As a result of this motion, the current exhibits oscillation.
This behavior is typical for this large oscillation case.
Because there is no current in the case of Fig. 4 (a), the oscillation amplitude is large compared to that in the small oscillation case.
If the dissipation is too strong, however, the piston-like motion is suppressed,
and there is only steady current, as described in Appendix B.

\begin{figure}
\begin{center}
  \includegraphics[width=70mm]{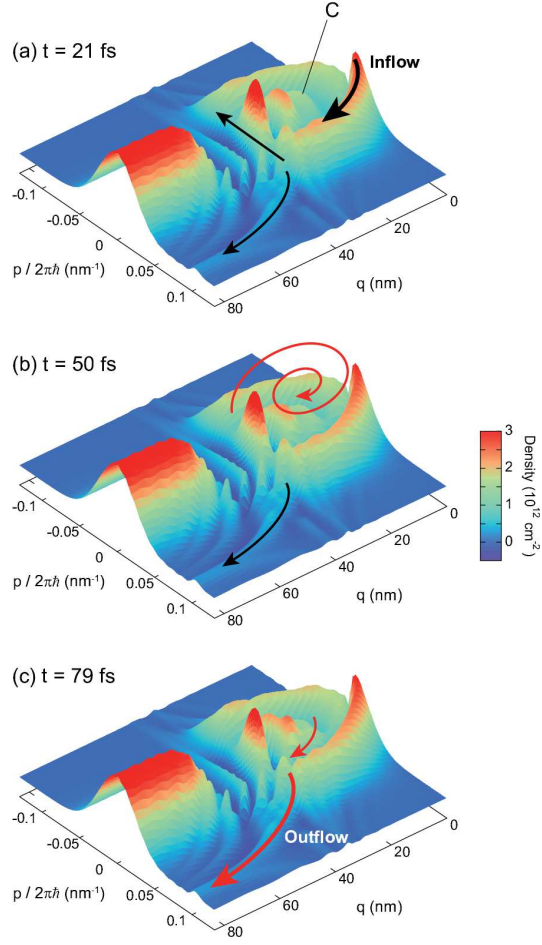}
\end{center}
\caption{\label{fig5}
Current oscillation with small amplitude observed in the upper part of the double-plateau structure
at the times marked on the green curve in figure 2 (ii) depicted as snapshots of the Wigner distribution.
(a) Current flows into the system from the emitter side of the boundary.
Then, a part of the current is scattered ($p_{in} \rightarrow -p_{in}$, where $p_{in}$ is the momentum of the inflow current) by the emitter side of the barrier.
The other part of the current flows to the collector side in the form of steady current through tunneling.
(b) The scattered electron flows in a tornado-like manner to peak C in the emitter basin due to dissipation.
(c) The shaking motion of the effective potential periodically accelerates the component at C to the tunneling state,
and the current is thus enhanced.}
\end{figure}

In figure 5, we display snapshots of the Wigner distribution for the case of small oscillation at the times marked in figure 2 (ii-b).
In figure 5 (a), while current flows into the system from the emitter side of the boundary, that part of it with energy larger than that of the tunneling state is scattered by the emitter side of the barrier.
The remaining current, i.e., that whose energy is closer to the energy of the tunneling state, flows to the collector side in the form of steady current, through tunneling.
In figure 5 (b), it is seen how the scattered electron moves back and forth in the emitter basin,
while losing energy due to dissipation.
As a result, the electron flows into peak C, exhibiting tornado-like motion.
Figure 5 (c) depicts shaking motion of the effective potential that periodically accelerates the electron distribution in peak C
to the tunneling state.
Through this effect, current flows to the collector side through the barrier.
Due to synchronization with this shaking motion, the current is enhanced periodically.
This tornado-like motion is typical for the small oscillation case.
Because there is a large contribution from steady current, the oscillation amplitude here is smaller than in the large oscillation case.

%========================================================================
\section{Conclusions\label{sec:concl}}

In summary, we investigated current oscillations in the plateau structures of the NDR region for three sizes of the contact regions
with a model that includes damping, employing the Caldeira-Leggett Hamiltonian.
We found two distinct types of current oscillations.
The first type is observed in the single plateau and in the lower part of the double-plateau structure.
It is characterized by a large oscillation amplitude and a single Fourier component.
The other type is observed in the upper part of double-plateau structure.
It is characterized by a small oscillation amplitude and two Fourier components.
An emitter basin that forms on the emitter side of the effective potential plays a key role in creating the current oscillation.
Eigenstate analysis indicates that the first type is caused by transitions between the ground tunneling state
and the adjacent excited state in the emitter basin,
while the second type is caused by transitions between the intermediate tunneling state and higher states.
Because the transition frequencies are large in the case of narrow emitter basin,
there is high frequency oscillation in the case of small contact regions for a fixed basin depth.
In Wigner space, these two types of oscillation are characterized by the two types of motion:
two-piston engine-like motion and tornado-like motion.
Dissipation plays an important role in the realization of current oscillation.
In order for the ground tunneling state to be populated in the case of large oscillation, there must be fairly strong dissipation,
whose strength is determined by the system-bath coupling and the noise correlation time.
If the dissipation is too strong, however, the current oscillation vanishes due to {damping}.
{The key to have non-trivial behaviors such as hysteresis, single/double plateaus and self-excited current oscillations is on the existence of the resonant tunneling states, the charge redistribution effects and the dissipation. The present results may be helpful to design nano-devices including a molecular junction system \cite{Nitzan2005}.}

Although many efforts have been made to improve Wigner transportation theory  \cite{Jacoboni2004,JiangCaiTsu2011}, the quantum Boltzmann equation \cite{HornbergerPR,HornbergerPRA}, and other formalisms  \cite{Feiginov2011,Knezevic2013} to study quantum dissipative dynamics in nano-devices, there are still a number of limitations and  many subtle problems on such formalisms, which deserve further attention.
On the basis of the reduced hierarchy equations of motion (HEOM) approach,
our investigation was carried out through highly accurate numerical calculations applied to a tunneling device system in a non-Markovian environment at finite temperature.
We have provided evidence of intrinsic bistability and self-excited current oscillations in the NDR region rigorously.
While the current oscillation experimentally observed in the RTD induced by the resonance with external circuit\cite{AsadaJJAP2008,SuzukiAPL2010}, however, these effects are not accounted for in our approach like many other theories based on the Boltzmann approach. To investigate the relation between the experimentally observed current oscillations and the existence of intrinsic current oscillation, further investigation is necessary.
Although the validity of the Caldeira-Leggett Hamiltonian in the description of
electron motion is not yet well established, we believe that the present results provide insight into the role of quantum mechanical phenomena in the type of system studied here. 

The present approach can be used to treat a strong system-bath coupling non-perturbatively.
In addition, any time-dependent external field can be added while taking into account the system-bath quantum coherence through the hierarchy elements.
Such features are ideal for studying SQUID rings \cite{Chen1986, Wellstood2008} and quantum ratchet systems \cite{Hanggi97,Hanggi09,KTJPCB2013}.

%////////////////////////////////////////////////////////////////////////
% Acknowledgments

\ack
YT is grateful for financial support received from the Humboldt Foundation and a Grant-in-Aid for Scientific Research (B2235006) from the Japan Society for the Promotion of Science.
AS acknowledges the research fellowship of Kyoto University.

\pagebreak

%////////////////////////////////////////////////////////////////////////
\appendix
\section{Effective potential and Wigner distribution in the case of decreasing bias}
\begin{figure}
\begin{center}
  \includegraphics[width=130mm]{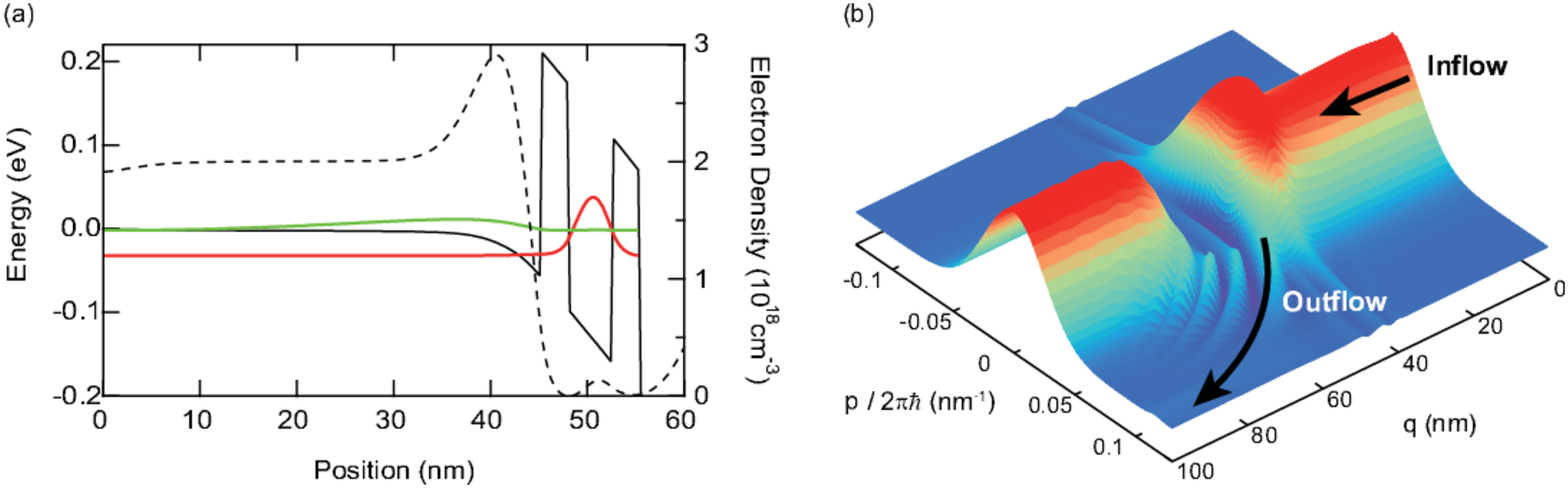}
\end{center}
\caption{\label{Layout3}
(a) The effective potential (black solid curve) and electron density (black dashed curve) in the steady-state in the case of decreasing bias. Here, the width of the contact region is 42.375 nm, and the bias voltage is 0.302 V. In contrast to the cases considered in figures 3 (i-b)-(iii'-b), in the case considered here, the emitter basin is very small. The red and green curves represent the first and second eigenstates.  (b) The steady-state Wigner distribution. The arrow indicates the direction of the steady current. 
}
\end{figure}
To understand the origin of hysteresis in the NDR region, we plot the effective potential, electron density
and energy eigenstates of the emitter basin and double-barrier well in the case of decreasing bias in figure  A1 (a).
In this case, the emitter basin is so shallow that there is no tunneling state between the emitter basin and the double-barrier well.
The existence of the peak near the barrier indicates that the second excited state is significantly populated.
The current arises through the transition from the second excited state to the ground quantum state through dissipation.
The Wigner distribution is plotted in figure A1 (b).
Due to dissipation, the distribution is in a steady state.
The peak near the barrier arises because the barrier impedes the flow.
The electron density then leaks to the collector side without oscillation through tunneling in the form of steady current.

\section{Strong coupling case}
To see the effect of dissipation, we determined the I-V characteristics for the case of a stronger system-bath coupling,
$\zeta=120.8$ GHz ($\zeta^{-1}=8.28$ ps).
The values of the other parameters are the same as in figure 2.
In contrast to the case considered in figure 2, in the present case current oscillation is not observed even in the plateau region.
Also, the upper plateau does not exist.
This is because the current oscillation decays quickly through the damping.
Because the heat bath is strongly coupled to the system, the eigenstate picture of the electron system itself,
as depicted in figure 3(b),
is of questionable validity.
As a result, the plateau is lost. 

\begin{figure}
\begin{center}
  \includegraphics[width=50mm]{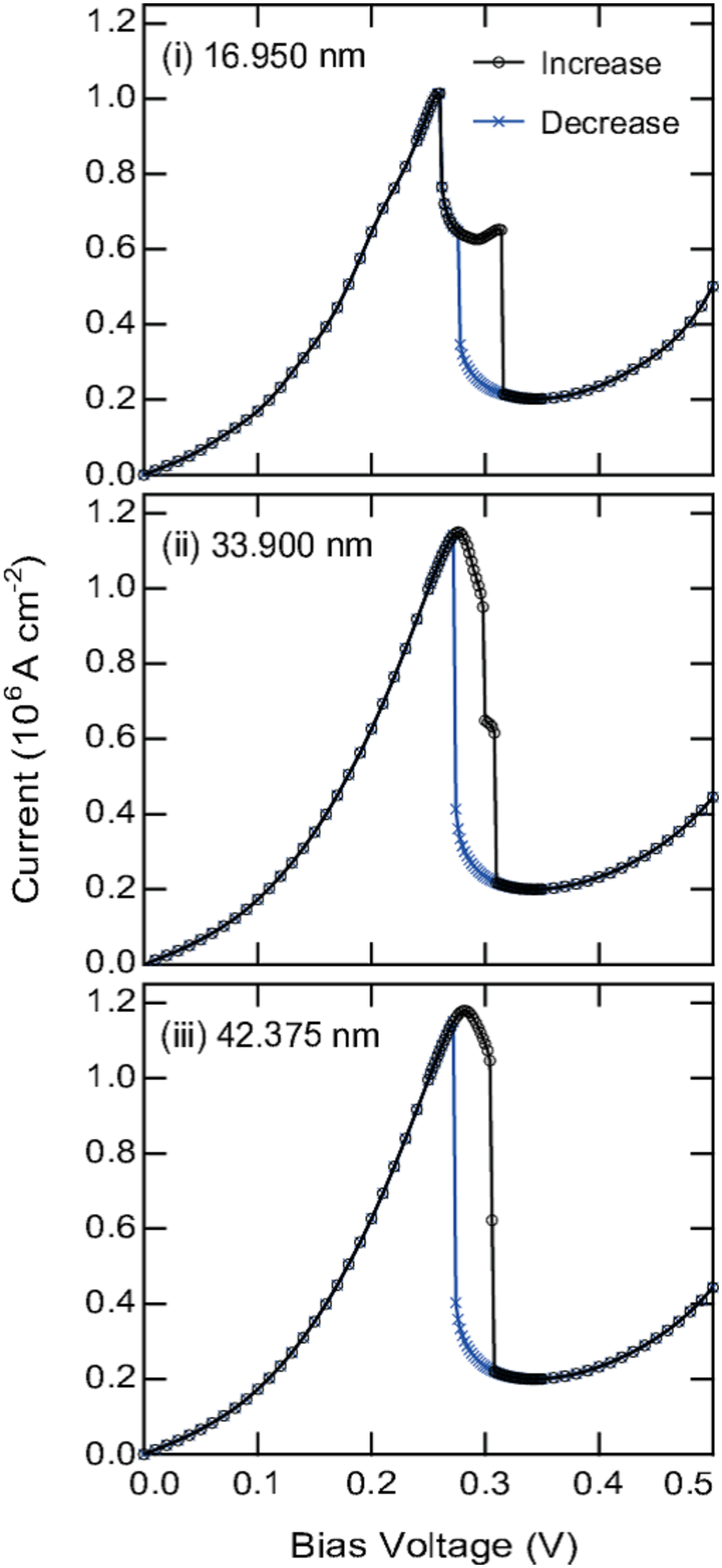}
\end{center}
\caption{\label{B1}
The I-V characteristics for the case of strong system-bath coupling, $\zeta=120.8$ GHz ($\zeta^{-1}=8.28$ ps).
The values of the other parameters are the same as in figure 2.
The black curve with the circles represents the case in which the bias is increasing,
and the blue curve with the $\times$s represents the case in which the bias is decreasing.
Comparing this figure with figure 2(a),
it is seen that the size of the NDR region is smaller and the plateau structure is less pronounced here
than in the weak-coupling case.
Current oscillation is not observed even in the plateau region. 
}
\end{figure}

\section{Comparison of Boltzmann results}
For the sake of comparison, we present the I-V characteristics calculated from the Boltzmann equation and the Poisson equation \cite{FerryPRB1989,BuotPRL1991,BuotIEEE1991,Plaummer1996,BuotJAP2000,Yoder2010}
for the same physical conditions as in figure 2.
The Boltzmann equation commonly used in the RTD problem is expressed as \cite{BuotPRL1991,BuotIEEE1991}
\begin{eqnarray}
  \frac{\partial}{\partial t} W(p, q; t)
  = - \hat{L}_\mathrm{qm}W(p, q; t) + \left( \frac{\partial  W(p, q; t)}{\partial t}   \right)_{\mathrm{coll}},
\end{eqnarray}
where $\hat{L}_\mathrm{qm}$ is the quantum Liouvillian defined by equations (\ref{Liouville1}) and (\ref{Liouville2}) and
%{$\left(\frac{\partial W}{\partial t} \right)_{\mathrm{coll}}$ is the collision operator.
%The rigorous treatment of collision operator is }
\begin{eqnarray}
 \left( \frac{\partial  W(p, q; t)}{\partial t}   \right)_{\mathrm{coll}}
=-\frac{1}{\tau} \left( W(p, q; t)- \frac{P(q; t)}{P_{\mathrm{eq}}(q)} W_{\mathrm{eq}}(p, q) \right)
\label{Collision}
\end{eqnarray}
is the modified collision operator under the relaxation time approximation. Here, $\tau$ is the relaxation time, $W_{\mathrm{eq}}(p, q)$ is the equilibrium Wigner distribution function, $P(q; t)=\int dp W(p,q;t) /(2\pi\hbar)$ is the density of the electron, and $P_{\mathrm{eq}}(q)$ is that of the equilibrium distribution, respectively. Because the collision term is determined by the Wigner distribution at time $t$, $W(p, q; t)$, and does not depend upon the previous history of distribution, this equation describes Markovian dynamics. Because the Boltzmann equation does not have a fluctuation term that is related to a dissipation term through the quantum version of the fluctuation-dissipation theorem, the equilibrium distribution is not an intrinsic state of the Boltzmann equation.
Moreover, we have to determine what the equilibrium distribution $W_{\mathrm{eq}}(p, q)$ is in an \textit{ad hoc} manner.

\begin{figure}
\begin{center}
  \includegraphics[width=50mm]{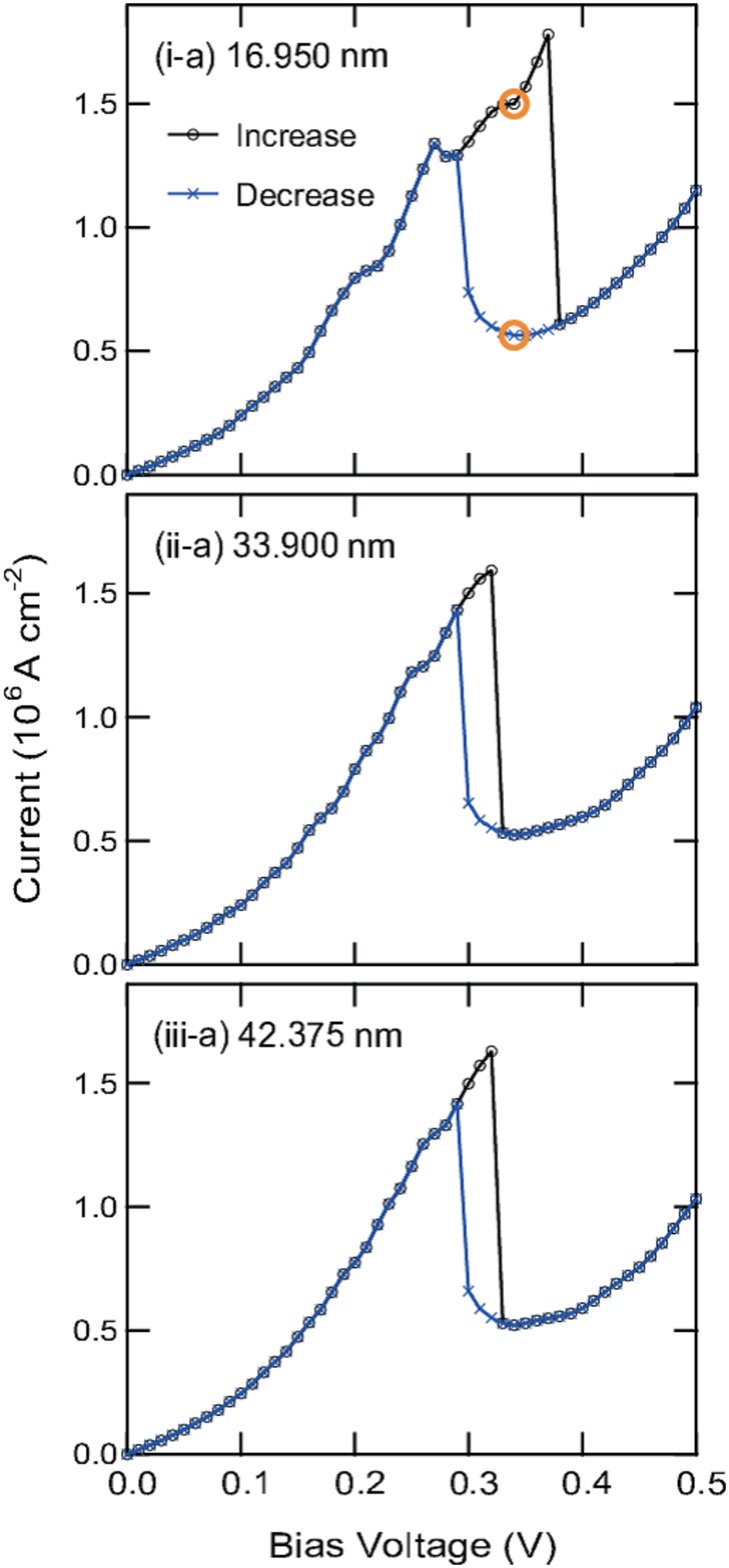}
\end{center}
\caption{\label{C1}
The I-V characteristics calculated from the Boltzmann equation and Poisson equation with $\tau=200$ ps. The values of the other parameters are the same as in figure 2. The black curve with circles represents the case in which the bias is increasing, and the blue curve with the $\times$s represents the case in which the bias is decreasing. The Wigner distributions at the voltage marked by the red circles in (i) are given in figure C2(a).
}
\end{figure}

We solve the Boltzmann equation for the effective potential calculated from equation (\ref{effectivepote}) and the Poisson equation (\ref{Poisson}) following the same procedure as in Sec. 4 with the same set of system parameters as in figure 2. 
Here, the equilibrium distribution, $W_{\mathrm{eq}}(p, q)$, is obtained from the quantum Liouville equation for the effective potential with the bias voltage zero \cite{BuotIEEE1991,BuotJAP2000}.
The boundary conditions is given by \cite{FrensleyPRB1987, JiangCaiTsu2011}
\begin{eqnarray}
 W(p,q=0 \ \mathrm{or}\ L) = \frac{m}{\pi\hbar^2\beta} \ln \left[1 + \exp\left( -\frac{\beta p^2}{2m}\right) \right].
\end{eqnarray}
In the Boltzmann equation approach, the time constant $\tau$ was estimated from other theory and Buot et al. set it to be $\tau=525$ fs at $T=77$K \cite{BuotIEEE1991}.
To compare with the HEOM result, here we solve the Boltzmann equation at $T=300$ K for various $\tau$ to find the case that exhibits similar I-V profiles as in figures 2(i)-2(iii). Note that the difference between the Fermi-Dirac and Bose-Einstein distributions is minor at this temperature.
The obtained results for $\tau=200$ fs are presented in figure C1.
In figures C1(i)-C1(iii), while we observed hysteresis, we could not find any plateau-like behavior and current oscillation in the NDR region as was shown in the HEOM calculations in figures 2(i)-2(iii).
The present results are also different from the result obtained from the Boltzmann equation at $T=77$K for $\tau=525$ fs, in which a single plateau behavior and current oscillations in the NDR region were observed \cite{BuotPRL1991, BuotJAP2000}. 

\begin{figure}
\begin{center}
  \includegraphics[width=120mm]{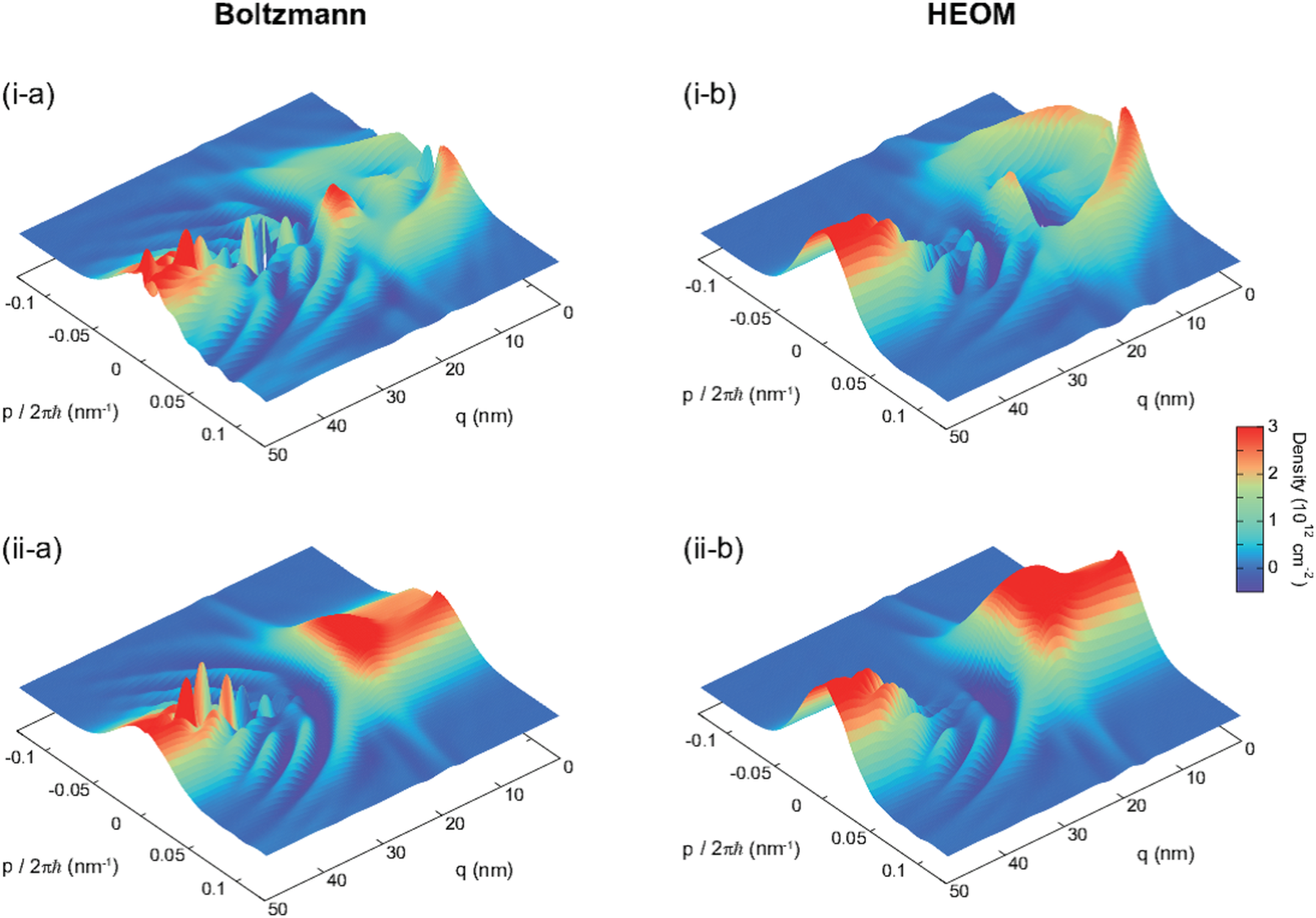}
\end{center}
\caption{\label{figC2}
The steady-state Wigner distribution for (i) the increasing bias case and (ii) the decreasing bias case calculated from (a) the Boltzmann equation and (b) the HEOM depicted in figure C1(i) and 2, respectively. The bias voltage of (i-a) and (ii-a) is 0.32 V (marked by the red circles in figure C1(i)), while that in (a-ii) and (b-ii) is 0.30 V. 
}
\end{figure}

To analyze the difference between the Boltzmann and HEOM results, we display snapshots of the steady-state Wigner distribution near the maximum and minimum of the I-V curves in figures C2(i) and 2(ii). We find that the Wigner distribution in the collector region is smooth in the HEOM case, while there are many small peaks disturbing the flow in the Boltzmann case. This difference arises because the equilibrium state in the Boltzmann approach is fixed even when the effective potential is changed from the original one due to the self-consistent calculations. As the result, the difference between the imposed equilibrium state and the true equilibrium state becomes large especially in the collector region, where the distribution is far from the assumed equilibrium distribution, $W_{\mathrm{eq}}(p, q)$, due to the scattering from the potential barriers.

As indicated in this appendix, dynamics described by the Boltzmann equation approach is different from the HEOM approach. This difference arises because the thermal equilibrium state of the Boltzmann equation is introduced as an assumption, while the thermal equilibrium state of the HEOM approach is an intrinsic state of the equation that is determined through the balance between the fluctuation term and dissipation term. 
This difference becomes significant for a system that exhibits hysteresis, since the equilibrium state of the system depends upon the pathway of process. This difference may also be significant if the system is driven by a time-dependent external field, in which the equilibrium state is not well-defined. 

\pagebreak
%======================================================================
% References
%.........................................................

%======================================================================
% @Tables

%.........................................................

%======================================================================
% @Figures

%.........................................................

\end{document}